\documentclass[a4paper,fleqn,usenatbib]{mnras}

\usepackage[T1]{fontenc}
\usepackage{ae,aecompl}

\usepackage{graphicx}	%
\usepackage{amsmath}	%
\usepackage{amssymb}	%

\newcommand{\Fig}[1]{Fig.~\ref{#1}}

\newcommand{\Sec}[1]{\S\ref{#1}}
\newcommand{\Tab}[1]{Table~\ref{#1}}

\newcommand{\remove}[1]{}

\newcommand{\jzd}[1]{}

\newcommand{\rewmgii}{\hbox{$EW_0^{\lambda 2796}$}}%
\newcommand{\rewsecondmgii}{\hbox{$EW_0^{\lambda 2803}$}}%

\newcommand{\incl}{i}

\newcommand{\NeV}{\textsc{[{\rm Ne}\kern 0.1em{\sc v}}]}
\newcommand{\NII}{\textsc{[{\rm N}\kern 0.1em{\sc ii}}]}
\newcommand{\OIII}{\textsc{[{\rm O}\kern 0.1em{\sc iii}}]}
\newcommand{\OII}{\textsc{[{\rm O}\kern 0.1em{\sc ii}}]}
\newcommand{\OI}{\textsc{{\rm O}\kern 0.1em{\sc i}}}
\newcommand{\MgI}{\textsc{{\rm Mg}\kern 0.1em{\sc i}}}
\newcommand{\MgII}{\textsc{{\rm Mg}\kern 0.1em{\sc ii}}}
\newcommand{\FeII}{\textsc{{\rm Fe}\kern 0.1em{\sc ii}}}
\newcommand{\MnII}{\textsc{{\rm Mn}\kern 0.1em{\sc ii}}}
\newcommand{\ZnII}{\textsc{{\rm Zn}\kern 0.1em{\sc ii}}}
\newcommand{\NaI}{\textsc{{\rm Na}\kern 0.1em{\sc i}}}

\newcommand{\HI}{\textsc{{\rm H}\kern 0.1em{\sc i}}}

\newcommand{\HII}{\textsc{{\rm H}\kern 0.1em{\sc ii}}}
\newcommand{\lya}{\textsc{{\rm Ly}\kern 0.1em$\alpha$}}
\newcommand{\Ly}{\textsc{{\rm Ly}\kern 0.1em$\alpha$}}
\newcommand{\Ha}{\textsc{{\rm H}\kern 0.1em$\alpha$}}
\newcommand{\Hb}{\textsc{{\rm H}\kern 0.1em$\beta$}}
\newcommand{\Hg}{\textsc{{\rm H}\kern 0.1em$\gamma$}}
\newcommand{\SII}{\hbox{[{\rm S}\kern 0.1em{\sc ii}}]}
\newcommand{\Ne}{\hbox{[{\rm Ne}\kern 0.1em{\sc v}}]}
\newcommand{\kms}{\hbox{km~s$^{-1}$}}
\newcommand{\kpc}{\hbox{kpc}}
\newcommand{\uerglf}{\mathrm{erg}\,\mathrm{s}^{-1}\,\mathrm{cm}^{-2}}

\newcommand{\mpy}{\mathrm{M}_{\odot}\,\mathrm{yr}^{-1}}

\newcommand{\myr}{\mathrm{Myr}}
\newcommand{\dex}{\mathrm{dex}}
\newcommand{\ue}{\mathrm{erg}\,\mathrm{s}^{-1}}
\newcommand{\um}{\mathrm{g}\,\mathrm{cm}\,\mathrm{s}^{-1}}
\newcommand{\ucoldens}{\mathrm{cm}^{-2}}

\newcommand{\vr}{v_\mathrm{r}}

\newcommand{\vcirc}{v_\mathrm{circ}}
\newcommand{\vvir}{v_\mathrm{vir}}

\newcommand{\vesc}{v_\mathrm{esc}}
\newcommand{\rvir}{r_\mathrm{vir}}

\newcommand{\vlos}{v_\mathrm{los}}

\newcommand{\vmax}{v_\mathrm{max}}
\newcommand{\ebv}{E(B-V)}

\newcommand{\Mvir}{M_\mathrm{vir}}
\newcommand{\Mstar}{M_*}

\newcommand{\hz}{h_\mathrm{z}}
\newcommand{\zabs}{z_\mathrm{abs}}
\newcommand{\msun}{\mathrm{M_{\odot}}}

\newcommand{\mdotout}{\dot{M}_\mathrm{out}}
\newcommand{\Edotout}{\dot{E}_\mathrm{out}}
\newcommand{\pdotout}{\dot{p}_\mathrm{out}}

\newcommand{\thetaout}{\theta_\mathrm{out}}
\newcommand{\thetain}{\theta_\mathrm{in}}
\newcommand{\vout}{v_\mathrm{out}}
\newcommand{\siggas}{\sigma_\mathrm{gas}}

\newcommand{\hr}{h_\mathrm{r}}

\newcommand{\gpk}{{\sc galpak3d}}{}
\newcommand{\mfl}{MEGAFLOW}{}
{}
\newcommand{\galfit}{{\sc galfit}}{}
{}
{}
\newcommand{\cgmpy}{{\sc cgmpy}}{}
{}
\newcommand{\ppxf}{{\sc ppxf}}{}
\newcommand{\coniecto}{{\sc coniecto}}{}

\newcommand{\backone}{\emph{back1}}{}
\newcommand{\backtwo}{\emph{back2}}{}
\newcommand{\main}{\emph{main}}{}

\newcommand{\compa}{\emph{A}}{}
\newcommand{\compb}{\emph{B}}{}
\newcommand{\compc}{\emph{C}}{}
\newcommand{\colcompa}{red}{}
\newcommand{\colcompb}{magenta}{}
\newcommand{\colcompc}{orange}{}

\newcommand{\fluxoiimain}{$(1.5{\scriptstyle \pm0.1})\times10^{-17}$}{} %
\newcommand{\mstarmain}{$9.8_{-0.0}^{+0.4}$}{} 
\newcommand{\ebvsedmain}{$0.00_{-0.00}^{+0.42}$}{}
\newcommand{\ebvmassmain}{$0.24_{-0.09}^{+0.12}$}{}
\newcommand{\agesedmain}{$9.5_{-0.3}^{+0.0}$}{} %
\newcommand{\sfroiiebvmassmain}{$0.5_{-0.2}^{+0.3}$}{} %
\newcommand{\ssfroiiebvmassmain}{0.07\pm0.06}{}
\newcommand{\sfrinstsedmain}{$0.3_{-0.0}^{+9.6}$}{} %
\newcommand{\tausedmain}{$8.7_{-0.1}^{+0.6}$}{} %
\newcommand{\vmaxmain}{$118{\scriptstyle \pm21}$}{} %
\newcommand{\sigmamain}{$38{\scriptstyle \pm15}$}{} %
\newcommand{\restbabsmain}{-19.6}

\newcommand{\vvirmain}{$107_{-30}^{+44}$}{} 
\newcommand{\rvirmain}{$120_{-34}^{+50}$}{} 
\newcommand{\mvirfromkinmain}{$11.5_{-0.4}^{+0.5}$}{}
\newcommand{\mvirfromstarsmain}{$11.6_{-0.1}^{+0.2}$}{} %

\newcommand{\deltamsmain}{$-0.6_{-0.6}^{+0.2}$}{} %

\newcommand{\vescqso}{261}{} %
\newcommand{\vescbackone}{287}{} %

\newcommand{\bmainarcsec}{2\arcsec.3} %
\newcommand{\bmainkpc}{16.8\,\kpc} %

\newcommand{\Rgalkpcbackone}{26\,\kpc} %
\newcommand{\Rgalrvirbackone}{0.21\,\rvir} %

\newcommand{\Rgalkpcqso}{51\,\kpc} %
\newcommand{\Rgalrvirqso}{0.42\,\rvir} %

\newcommand{\vuvescompa}{-49} %
\newcommand{\vuvescompb}{10} %
\newcommand{\vuvescompc}{100} %

\newcommand{\vrforcompAkms}{-40\,\kms} %
\newcommand{\vrforcompAvvir}{-0.4\,\vvir} %

\newcommand{\vmusequasar}{25.8} %
\newcommand{\evmusequasar}{0.4} %
\newcommand{\vmusebackone}{110} %
\newcommand{\evmusebackone}{17}
\newcommand{\veldiffmuse}{84} %
\newcommand{\eveldiffmuse}{17}

\newcommand{\bkpcquasarmm}{16.8} %
\newcommand{\bkpcbackonemm}{8.8} %
\newcommand{\bkpcbacktwomm}{20.5} %

\newcommand{\alphaquasarmm}{81} %
\newcommand{\alphabackonemm}{-109} %
\newcommand{\alphabacktwomm}{115} %

\newcommand{\zrefprecise}{0.70344} %

\defcitealias{Zabl:2019a}{paper\,II}

\newcommand{\papone}{paper\,I}{}
\newcommand{\paptwo}{paper\,II}{}
\newcommand{\papthree}{paper\,III}{}

\title[Tomography of a galactic wind]{MusE GAs FLOw and Wind (MEGAFLOW) IV: A two
  sightline tomography of a galactic wind}

\author[J. Zabl et al.]{
Johannes Zabl,$^{1}$\thanks{E-mail: johannes.zabl@univ-lyon1.fr}
Nicolas F. Bouch\'e,$^{1}$
Ilane Schroetter,$^{2}$
Martin Wendt,$^{3}$
\newauthor
Thierry Contini,$^{4}$
Joop Schaye,$^{5}$
Raffaella A. Marino,$^{6}$
Sowgat Muzahid,$^{5}$
\newauthor
Gabriele Pezzulli,$^{6}$
Anne Verhamme,$^{7}$
Lutz Wisotzki$^{8}$
\\
$^{1}$ Univ Lyon, Univ Lyon1, Ens de Lyon, CNRS, Centre de Recherche Astrophysique de Lyon UMR5574, F-69230 Saint-Genis-Laval, France\\
$^{2}$ GEPI, Observatoire de Paris, CNRS-UMR8111, PSL Research University, Univ. Paris Diderot, 5 place Jules Janssen, 92195 Meudon, France \\
$^{3}$ Institut f\"ur Physik und Astronomie, Universit\"at Potsdam, Karl-Liebknecht-Str. 24/25, 14476 Golm, Germany\\
$^{4}$ Institut de Recherche en Astrophysique et Plan\'etologie (IRAP), Universit\'e de Toulouse, CNRS, UPS, F-31400 Toulouse, France \\
$^{5}$  Leiden Observatory, Leiden University, PO Box 9513, NL-2300 RA Leiden, the Netherlands\\
$^{6}$ Department of Physics, ETH Z\"urich,Wolfgang-Pauli-Strasse 27, 8093 Z\"urich, Switzerland \\
$^{7}$ Observatoire de Gen\'eve, Université de Gen\'eve, 51 Ch. des Maillettes, 1290 Versoix, Switzerland \\
$^{8}$ Leibniz-Institut f\"ur Astrophysik Potsdam (AIP), An der Sternwarte 16, 14482 Potsdam, Germany \\
}

\date{Accepted XXX. Received YYY; in original form ZZZ}

\pubyear{2015}

\begin{document}

\label{firstpage}
\pagerange{\pageref{firstpage}--\pageref{lastpage}}
\maketitle

\begin{abstract}
Galactic outflows are thought to eject baryons back out to the
circum-galactic medium (CGM).
Studies based on metal absorption lines  (\MgII{} in particular) in the spectra of background quasars indicate that the
gas is ejected anisotropically, with galactic winds likely leaving the host
in a bi-conical flow perpendicular to the galaxy disk.
 In this paper, we present a detailed analysis of an outflow from a $z=0.7$
``green-valley" galaxy ($\log(\Mstar/M_\odot)=9.9$; $\mbox{SFR}=0.5\,\mpy$) probed by two background sources part of the MUSE Gas Flow and Wind (MEGAFLOW) survey.
Thanks to a fortuitous configuration with a background quasar (SDSSJ1358$+$1145)
and a bright background galaxy at $z=1.4$, both at impact parameters of  $\approx15\,\kpc$, we can -- for the first time -- probe
\emph{both} the receding and approaching components of a putative galactic outflow around a distant galaxy. 
We measure a significant velocity shift between the \MgII{} absorption from the
two sightlines ($84\pm17\,\kms$), which is consistent with the expectation from our simple fiducial
wind model, possibly combined with an extended disk contribution.
    
\end{abstract}


\section{Introduction}

Galaxies are surrounded by a complex multi-phase medium, the circumgalactic medium (CGM; \citealt[]{Tumlinson:2017a} for a recent review).
Accretion from this CGM onto galaxies and winds from the galaxies into the CGM are believed to be key ingredients in regulating the evolution of galaxies.

The detailed study of absorption features detected in bright background sources is one of the main observational tools helpful in characterizing the physical properties and kinematics of the CGM gas.
Among various transitions, the $\MgII{}\lambda\lambda 2797,2803$ doublet is
an especially useful tracer of the cool, photo-ionized component of the CGM
($T\approx10^{4-5} \mathrm{K}$; e.g.,~\citealt{Bergeron:1986a}).
Its strength, easy identifiability as a doublet, and convenient rest-frame wavelength have allowed the collection of large statistical samples of \MgII{} absorbers \citep[e.g.][]{Lanzetta:1998a, Steidel:1992a, Nestor:2005a, Zhu:2013a} at redshifts $0.1 \lesssim z \lesssim 2.5$.
Follow-up observations of the fields surrounding the absorbers have identified galaxies associated to the absorbers and, hence, clearly established that the \MgII{} absorbing gas is found in the haloes of galaxies \citep[e.g.][]{Bergeron:1988a, Bergeron:1991a, Steidel:1995a, Steidel:2002a, Nielsen:2013a, Nielsen:2013b}.

Subsequently, large observational efforts have been put into mapping the spatial distribution and kinematics of the \MgII{} absorbing gas w.r.t. the galaxies in whose haloes the gas resides. 
The major result from these studies is that the $\MgII{}$ absorbing gas is not
isotropically distributed around the galaxies \citep[e.g.][]{Bordoloi:2011a,
  Bouche:2012a, Kacprzak:2012a, Lan:2014a, Lan:2018a, Zabl:2019a, Martin:2019a, Schroetter:2019a}.
Instead, the observations support a two-component geometry: a bi-conical outflow perpendicular to the galaxy disk and an extended gas disk approximately co-planar with the stellar disk.
This allows to split the \MgII{} absorber sightlines into an outflow and a disk
sub-sample, which can be used to study the kinematics of the outflows
\citep[e.g.][]{Bouche:2012a, Kacprzak:2014a, Muzahid:2015a, Schroetter:2015a,
  Schroetter:2016a, Schroetter:2019a, Rahmani:2018b, Martin:2019a} and the extended gas accretion disks \citep[e.g.][]{Steidel:2002a, Chen:2005a, Kacprzak:2010a, Kacprzak:2011a, Bouche:2013a, Bouche:2016a, Ho:2017a, Ho:2019a, Rahmani:2018a, Zabl:2019a}, respectively.

The aforementioned results have been obtained statistically by collecting single sightlines around \emph{many galaxies}.
A step forward would be to directly map the geometry of the CGM around \emph{individual galaxies.}
Such ``tomography'' requires multiple or very extended bright background sources behind the CGM of an individual galaxy. 

Taking advantage of the comparably large extent that galaxies in the local Universe span on the sky, \citet{Bowen:2016a} have used four different background quasars to firmly conclude for an individual galaxy that the absorbing gas is distributed in an extended gas disk.
However, having multiple sufficiently bright background galaxies covering the halo of a single galaxy is rare, especially at high redshift where the virial radius corresponds to a fraction of an arcminute. 

The few studies beyond the local Universe were either using quasars by chance aligned close to each other \citep[e.g.][]{DOdorico:1998a,Crighton:2010a,Muzahid:2014a}, multiple imaged lensed-quasar pairs
\citep[e.g.][]{Rauch:1999a, Lopez:1999a, Lopez:2007a, Ellison:2004a, Rubin:2018b}, or extended galaxies \citep[e.g.][]{Peroux:2018a,Lopez:2018a, Lopez:2019a}. The main focus of these studies was to 
characterize the coherence scale of the absorbing gas.

In this paper, we present a tomographic study of the CGM around a $z=0.70$
galaxy surrounded by two bright background sightlines which was discovered in
the MUSE Gas FLow and Wind (\mfl{}) survey (\citealt{Schroetter:2016a}
-\papone{}-; \citealt{Zabl:2019a} -\paptwo{}-; \citealt{Schroetter:2019a} -\papthree{}-).
 This survey consists of 79 strong \MgII{} absorbers towards 22 quasar sightlines which have been selected to have (at least) three $\MgII{}$ absorbers with rest-frame equivalent widths $\rewmgii>0.3\,\text{\AA}$ and $0.4 < \zabs < 1.5$.

The paper is organized as follows. We present our observations in \Sec{sec:observations},  the galaxies and absorption sightlines in the field in \Sec{sec:result}, a  model for the CGM in \Sec{sec:model}. We compare our CGM model to our data  and  discuss our results in \Sec{sec:discussion}. Finally, we present our conclusions  in \Sec{sec:conclusions}.
Throughout, we use a 737 cosmology ($H_0=70$~\kms, $\Omega_{\rm m}=0.3$, and $\Omega_{\Lambda} = 0.7$) and we state all distances as 'proper' (physical) distances. A \citet{Chabrier:2003a} stellar Initial Mass Function (IMF) is assumed.
We refer to the $\OII\,\lambda\lambda3727,3729$ doublet simply as \OII{}.
All wavelengths and redshifts are in vacuum and are corrected to a heliocentric velocity standard.

\section{Observations}
\label{sec:observations}

\subsection{MUSE data}

We observed the field around the quasar SDSSJ1358+1145 with MUSE (Multi Unit Spectroscopic Explorer;
\citealt{Bacon:2006a,Bacon:2010a}) for a total integration time
of 3.11\,hr.
The first four exposures (4x1500\,s=1.67\,hr; 2016-04-09), which
constitute the data used in papers II\&III, were taken with the nominal
wide field mode without adaptive optics (AO) (WFM-NOAO-N), as MUSE's AO system was not yet available at the time.
After identifying the science case of the present work, we realized a potential benefit from using MUSE's extended mode for subsequent observations of the field. Therefore, we completed the observations in extended wide field mode, while additonally taking advantange of the available AO (4x1300\,s=1.44\,hr; 2018-03-14; WFM-AO-E).
Extended mode increases the blue wavelength coverage from $4750\,\textnormal{\AA}$ to $4600\,\textnormal{\AA}$ with the trade-off of some second order contamination at wavelengths $\gtrsim 8000\,\textnormal{\AA}$.
The extra coverage helps to better constrain the continuum around $\MgII\,\lambda2796$ at $z=0.704$, the redshift of the foreground galaxy whose CGM we study in this work.

We reduced the data identically to \paptwo{}, except that we were using DRSv2.4 \citep{WeilbacherP:2012a,WeilbacherP:2014a,WeilbacherP:2016a}, which allows for the reduction of the AO data.
The combined AO and non-AO data have a point source Moffat full width at half maximum (FWHM) of $0\arcsec.55$ at $7050\,\text{\AA}$. Using the depth estimator from \paptwo, this exposure time (3.11\,hr) and this seeing results in an \OII{} point source detection limit of $2.7\times10^{-18}\,\uerglf$. \footnote{The estimate is for $\approx 7000\,\textnormal{\AA}$. The detection limits are higher at shorter and longer wavelengths (see e.g. \citealt{BaconR:2017a}).}

\subsection{UVES data}

We observed the quasar SDSSJ1358+1145 with the VLT high-resolution spectrograph UVES (Ultraviolet and Visual
Echelle Spectrograph; \citealt{Dekker:2000a}) for a total integration time of $2966\,\mathrm{s}$ in the night of 2016-04-07. 
Further details about observation, reduction, and continuum normalisation are given in \citetalias{Zabl:2019a}.

 \section{Result}
 \label{sec:result}

\subsection{Identification of background sightlines}

The main galaxy at $z=0.704$ (\main{}) was discovered through association with an $\rewmgii=2.5\,\textnormal{\AA}$ \MgII\ absorber towards the quasar SDSSJ1358+1145 from \mfl{} at an impact parameter of $b=\bmainarcsec$ ($\bmainkpc$).

This quasar sightline is particularly interesting as it contains two additional very strong \MgII{} absorbers with a rest-frame equivalent width $\rewmgii=1.8$ and $2.6\,\textnormal{\AA}$ at redshifts $\zabs=0.81$ and 1.42, respectively.
The galaxy counterparts of the $\zabs=0.81$ and $\zabs=1.42$ absorbers have been described in \papthree{} (wind sample) and \paptwo{} (accretion sample), respectively. 
They are galaxies with $\log(\Mstar/\msun)$ of 9.3 and 9.9, and are at relatively small impact parameters of $1\arcsec.6$ and $3\arcsec.6$ from the quasar, as also expected from the known \MgII{} EW--impact parameter anti-correlation \citep[e.g.][]{Lanzetta:1990a, Bouche:2006a, Kacprzak:2011b, Chen:2010a, Nielsen:2013b}.
We refer to these galaxies in the following as \backtwo{} and \backone{}, respectively.

Thus, together with the quasar, the \main{} ($z=0.704$) galaxy has potentially
three background sightlines (quasar, \backone{, and \backtwo}) that can probe
the CGM kinematics. In addition to the quasar, \backone\ is a useful background
source, as it has a very bright UV continuum.\footnote{The full spectral energy
distribution (SED) of the z=1.42 \backone{} galaxy is shown in the Supplementary
Appendix of \paptwo{}. The galaxy has a $M_{2800\,\textnormal{\AA}}$ absolute
total magnitude of -20.8, which is slightly brighter than the characteristic
Schechter magnitude at its redshift \citep{Dahlen:2007}.} \backtwo{} is not a
useful background source, due to intractable contamination from the close-by
quasar. The orientation of all three sightlines w.r.t. the $z=0.704$ galaxy is shown
in Fig.~\ref{fig:kinematics_model}(A) and listed in \Tab{tab:orientation}.
The listed uncertainties are resulting from the uncertainties on position
angle and centroid of the \main{} galaxy (cf. \Sec{sec:gal_properties} and
Appendix~\ref{app:uncert_incl_posangl}).

\subsection{The main galaxy's properties}
\label{sec:gal_properties}
\begin{figure}
    \includegraphics[width=1.0\columnwidth]{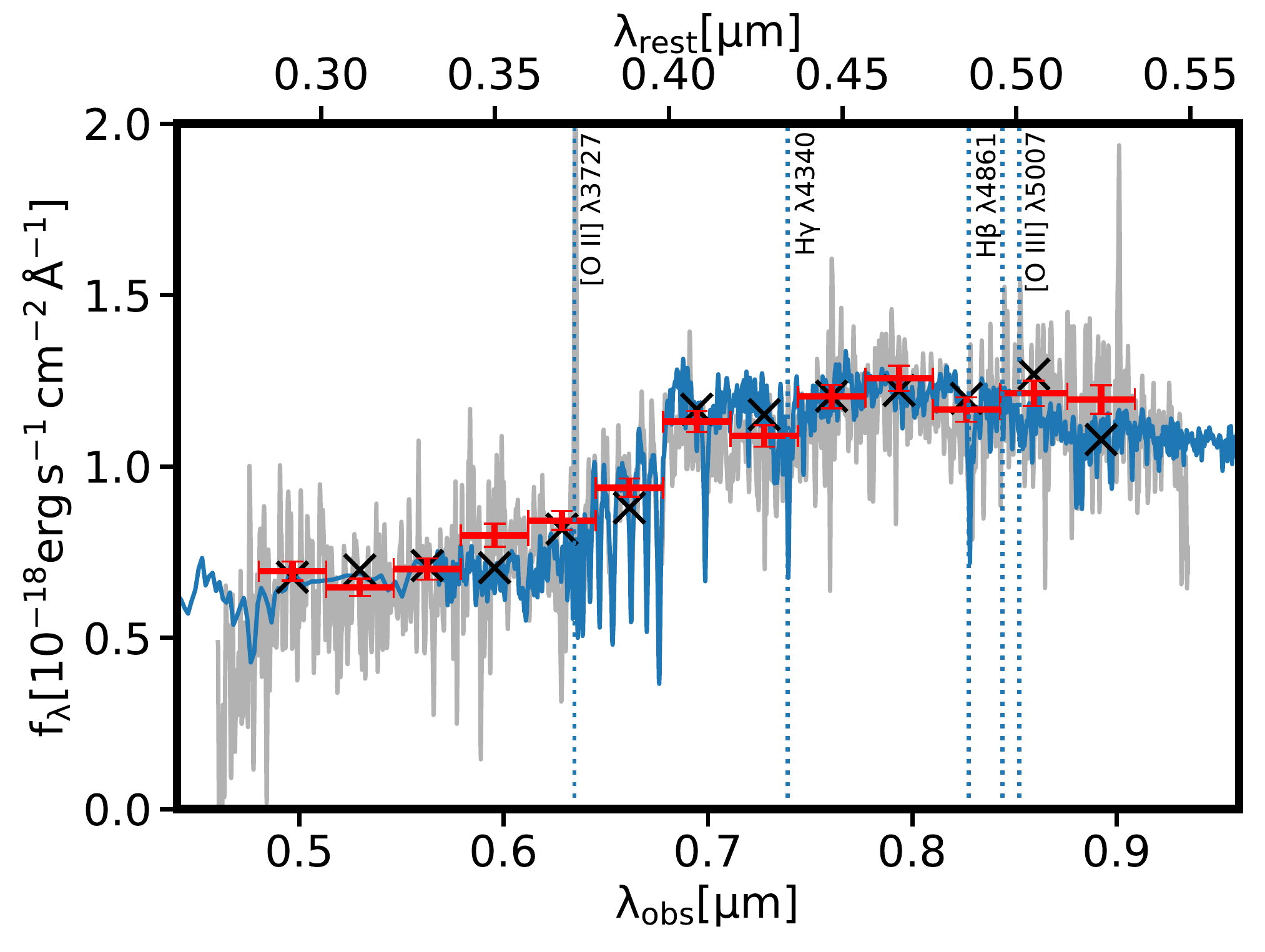}
    \caption{\label{fig:sed}
    Best-fit SED model (blue) for the main foreground galaxy compared to the
    observed spectrum (grey, smoothed with $FWHM=8.8\,\textnormal{\AA}$ Gaussian).
     The fit was done using pseudo-photometry for 13
    medium-band filters created from the spectrum itself.
     The red errorbars indicate the filter-averaged flux densities in these filters, with the horizontal bars indicating the width of the filters. 
    The black crosses show the flux-densities in the same filters as obtained from the best-fit SED.
    While the SED fitting was done including emission lines and the shown model medium band flux-densities include this contribution, 
    the best-fit SED is shown without the emission to avoid visual confusion with the actual emission lines.
    }
\end{figure}

The spectrum of the main $z=0.704$ galaxy is shown in Fig. \ref{fig:sed}. The
galaxy shows visibly weaker line emission than is typical for star-forming
galaxies on the star forming ``main sequence'' (MS) at this redshift
\citep[e.g.][]{Speagle:2014, Boogaard:2018a}. Quantitatively, we found the
galaxy to have a stellar mass of $\log(M_*/\msun)=$\mstarmain{} and a star
formation rate ($\mbox{SFR}$) of \sfroiiebvmassmain$\,\mpy$.

 The corresponding specific SFR ($\mbox{sSFR}=\ssfroiiebvmassmain\,\mathrm{Gyr}^{-1}$) is \deltamsmain$\,\dex$ (or $\approx 1.5\sigma$) below the MS prediction for $z=0.70$ \citep{Boogaard:2018a}. This means our galaxy is similar to 'green valley' galaxies. 
 
We determined the stellar mass and $\mbox{SFR}$ as in \citetalias{Zabl:2019a}.
In short, we estimated $M_*$ from SED fitting using our custom code \coniecto{}
(see also \citealt{Zabl:2016a}) on 13 pseudo-medium band filters created from the
MUSE spectrum.\footnote{Different from \paptwo{}, we assumed a delayed $\tau$ star
  formation history (SFH) ($\mbox{SFR}\propto t \times exp(-t/\tau)$, with $t$ being the elapsed cosmic time since the galaxy started forming stars.}
Other values obtained from the SED fit are listed in \Tab{tab:prop_main_gal}.
The (instantaneous) $\mbox{SFR}$ was determined starting from the measured \OII{} flux, correcting it for extinction using the \citet{Calzetti:2000a} law with the strength of the extinction estimated from the $M_*-\ebv$ relation of \citet{Garn:2010a}, and converted to a SFR using the \citet{Kewley:2004a} relation.

We estimated the \OII{} flux from a fit to the \OII{} morpho-kinematics using
the 3D fitting tool \gpk{} \citep{Bouche:2015a}. This fit provided us also with
a best-fit estimate of the kinematics (see \Tab{tab:prop_main_gal}). The steps
involved in the \gpk{} fitting were again identical to those described in
\paptwo{}. However, as the \OII{} flux is low for this galaxy, it was not
possible to robustly measure the kinematics and morphology (inclination in
particular) based on \OII{} alone.\footnote{This is the reason why the galaxy
was not part of the sample in \papthree{}.} Thus, we decided to constrain the
inclination, $\incl$, using a continuum map in a pseudo r-band image created
from the MUSE cube. We determined the galaxy morphology, including $\incl$ and
position angle, $PA$, from this continuum map using \galfit{}
\citep{Peng:2010a}. Further, we used the appropriate PSF for the r-band as determined
from the quasar. The fit was complicated by systematic residuals from
the close-by quasar. Nevertheless, we could obtain a robust estimate of
$\incl=71\pm5\deg$ and $PA=37\,\pm 8\deg$. Details about the fit and the
method to estimate the uncertainties are given in
Appendix~\ref{app:uncert_incl_posangl}. Finally, we fit the \OII{} kinematics
with \gpk{} using $\incl$ and the $PA$ as obtained from the
continuum ($i=71\deg$, $PA=37\deg$).

\subsection{Absorption in CGM of the main galaxy}

The CGM around the $z=0.704$ main galaxy can be probed in absorption at multiple locations using the spectra of the background quasar and the \backone{} galaxy.
While high spectral resolution spectroscopy is only available for the quasar,  we can use the MUSE data cube to probe \MgII{} absorption with the same spectral resolution in both sightlines.

\subsubsection{\MgII{} absorption at the resolution of MUSE}
\label{sec:mgii_abs_muse}

\begin{figure*}
  \includegraphics[width=1.0\textwidth]{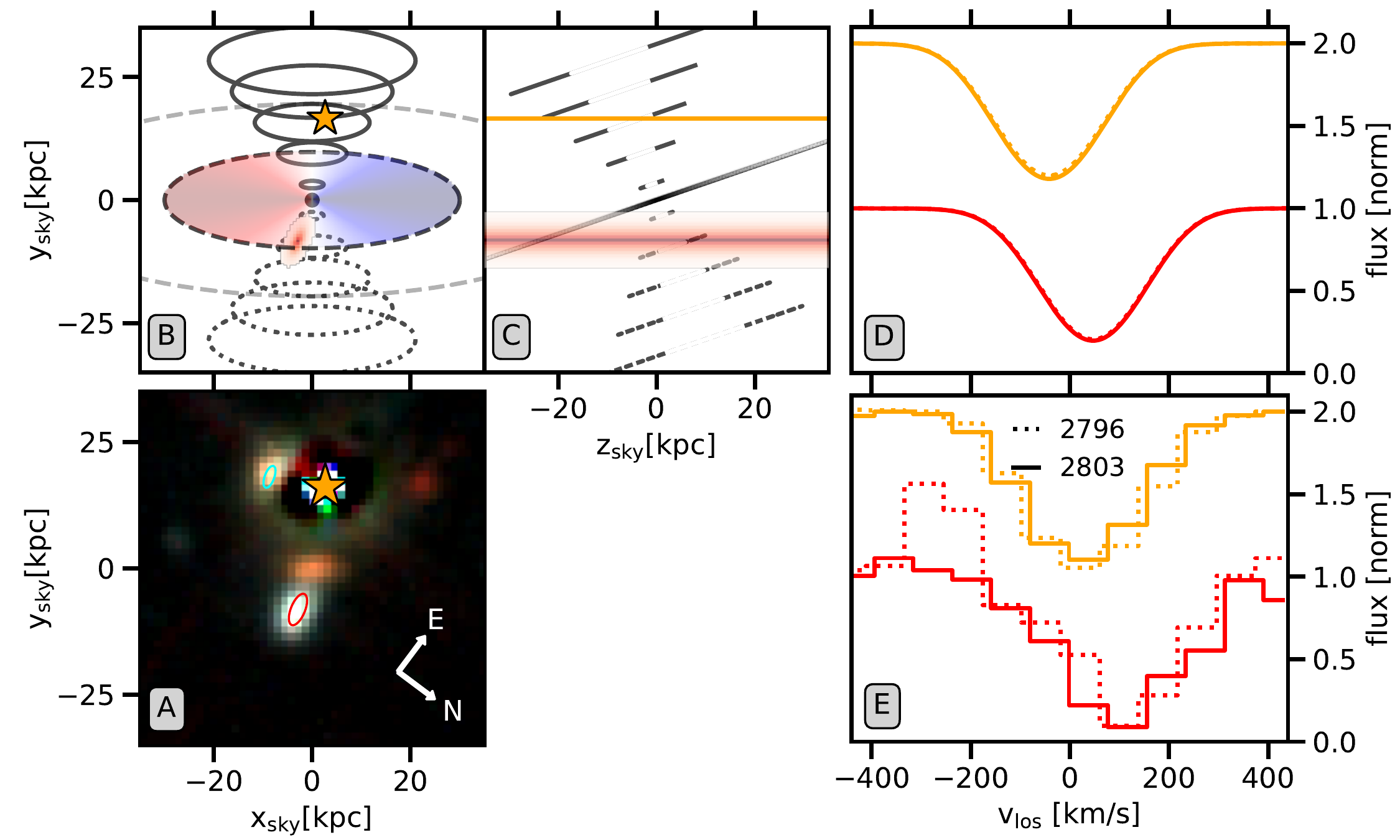}

  \caption{\label{fig:kinematics_model}
    Comparison between data and model for the \MgII{} absorption seen in the MUSE spectrum for two sightlines through the CGM of the \main{} galaxy. \textbf{A:} $9\arcsec{}.8\times9\arcsec{}.8$ field, corresponding to $70\ \kpc\times70\,\kpc$ at $z=0.70$, around the \main{} foreground galaxy  shown as color image with pseudo z',r',V broad-band MUSE images in the red, green, and blue channel, respectively. The quasar was subtracted, but residuals are visible.
    The \main{} foreground galaxy (center) is surrounded by three bright background sources: the \emph{quasar} towards the top (orange star), the bright galaxy towards the bottom (\backone{}; red ellipse), and the second galaxy close to the quasar (\backtwo; cyan ellipse). 
   \textbf{B:} View of the assumed CGM model (see \Sec{sec:model}) on the sky
   plane. The approaching outflow cone is indicated as solid concentric circles,
   while the receding outflow cone is indicated by dotted circles. For the
   extended gas disk, the rotation line-of-sight velocity field is overlaid. The
   orientation is identical to panel A and the positions of the quasar and 
   \backone{} are indicated by the orange dot and the red surface-brightness ellipse, respectively.
   \textbf{C:} Geometry of the same model as in B, but here with the line-of-sight direction on the x-axis.
    The point-source sightline for the quasar (orange) and the extended sightline for 
   \backone{} (red) are indicated. 
   The cone is hollow in the inner part.
   \textbf{D:} \MgII{$\,\lambda\lambda2796,2803$ line-of-sight kinematics simulated at the resolution of MUSE based on the model shown in panels B\&C and described in \Sec{sec:discussion:disk_and_wind} for the quasar (orange; offset by +1) and \backone (red) sightlines,
   respectively. The model parameters are listed as 'disk + wind' model in \Tab{table:models:params}.}
   \textbf{E:} \MgII{} line-of-sight kinematics measured with MUSE in the background quasar (orange; offset by +1) and galaxy
   spectra (red), respectively. Both the $2796\,\textnormal{\AA}$ and $2803\,
   \textnormal{\AA}$ lines of the \MgII{} doublet are shown (dotted/solid). The zero-velocity corresponds to the systemic redshift of the \main{} galaxy ($z=\zrefprecise$) as measured from the \OII{} emission.
          }
\end{figure*}

\MgII{} is the strongest among the CGM metal  absorption lines covered by the MUSE data at this redshift and hence the most useful to probe the CGM with low signal-to-noise (S/N) background galaxy sightlines. 
We show in panel E of \Fig{fig:kinematics_model} the observed z=0.704 \MgII{}
absorption  both for the quasar (orange) and the \backone{} (red) sightlines ($\MgII \,\lambda 2796$ -dotted-, $\MgII\,\lambda 2803$ -solid-).
The figure shows that \MgII{} absorption  is not only visible in the quasar sightline ($\rewmgii=2.7\,\textnormal{\AA}$), as per selection, but also in the $\backone$ sightline ($\rewmgii=2.0\,\textnormal{\AA}$).

Despite the moderate spectral resolution ($190\,\kms$ at $4700\,\textnormal{\AA}$), the absorption profiles encode interesting information. 
First, a velocity shift is clearly visible between the two sightlines. The absorption in the \backone{} galaxy sightline is redshifted w.r.t. that in the quasar sightline by $\veldiffmuse\pm\eveldiffmuse\,\kms$, with the absorption in the two sightlines centred at $\vmusebackone\pm\evmusebackone\,\kms$ and $\vmusequasar\pm\evmusequasar\,\kms$, respectively. We obtained these velocity measurements by simultaneously fitting both components of the \MgII{} doublet with Gaussians.
Second, we measured a $\rewmgii$/$\rewsecondmgii$ ratio close to one in both sightlines. This means the \MgII{} absorption is strongly saturated.\footnote{The $\rewmgii$/$\rewsecondmgii$ ratio for optically thin absorption  is 2:1.}
Third, we find that the flux reaches almost zero at peak absorption. For both sightlines, this means, when accounting for the resolution of MUSE, that the \MgII{} absorption is spread over a large velocity range. For the extended galaxy sightline (\backone), this further means that the \MgII{} coverage must be complete over the extent of the aperture from which we have extracted the background spectrum. The non-circular extraction aperture, which was chosen to optimize the S/N, included 29 spatial pixels corresponding to an area of $1.2\,\mathrm{arcsec}^2$.

\subsubsection{Absorption at the resolution of UVES}
\label{sec:uves_abs}

\begin{figure}
  \centering
    \includegraphics[width=0.76\columnwidth]{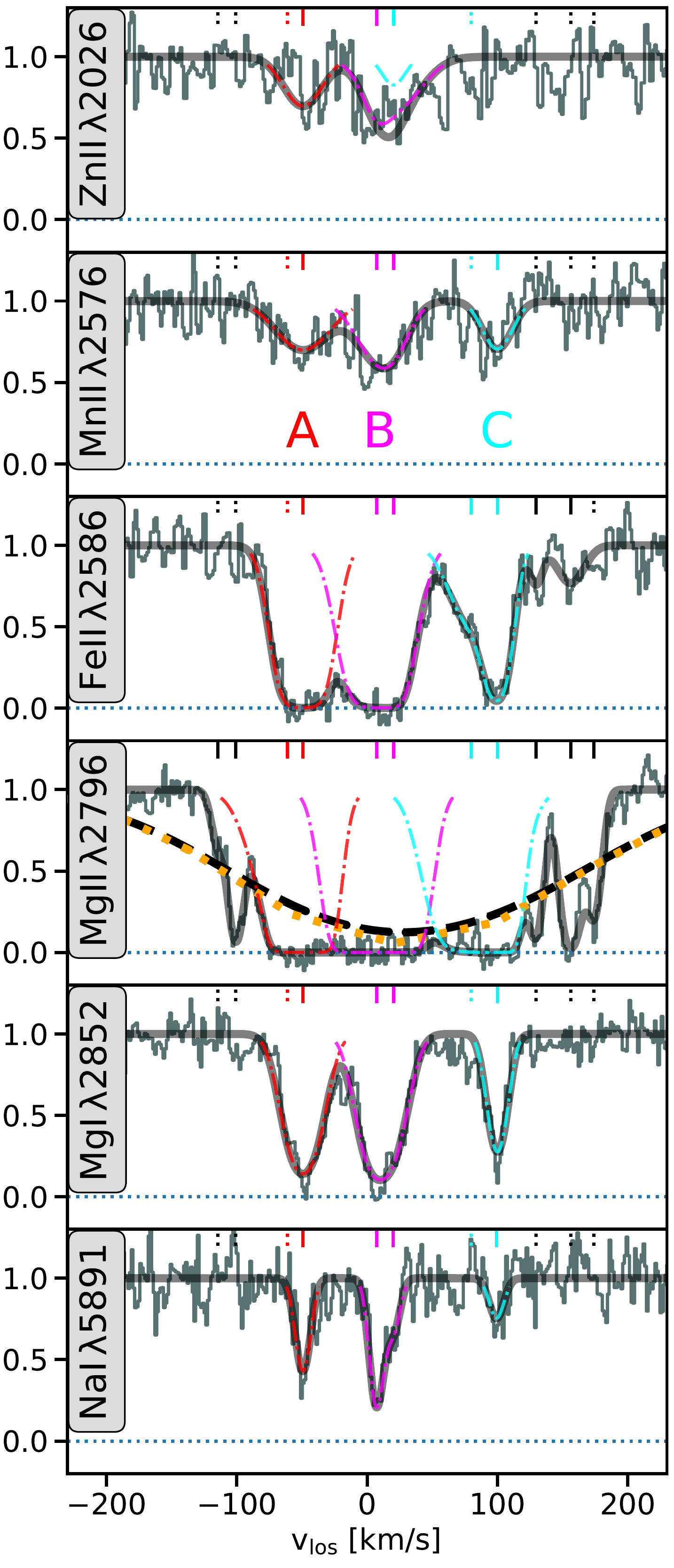}
    \caption{
      \label{fig:uves_spec}
      Absorption in the quasar sightline at the redshift of \emph{main} measured with the high spectral resolution VLT/UVES data.  {\bf Panels 1-6 from the top:} The
      observed absorption is shown for multiple species, with a multi-component model fit
      (thick grey line) overplotted over the data. The velocity components
      considered in the fit are indicated as little bars near the top of the
      panels, where a dotted bar indicates that the component was not used for
      the specific line.
      Three main kinematic components, A, B, C, can be clearly identified from
      the unsaturated lines. The contribution of the three components, as
      measured from the multi-component fit, is shown by different colors.
      For comparison, the panel for $\MgII{}\,\lambda 2796$ also shows the MUSE spectrum (orange dotted; identical to panel E in \Fig{fig:kinematics_model}) and the UVES spectrum artificially degraded to the resolution of MUSE (black dashed).
       }
\end{figure}

In the previous section, we compared \MgII{} absorption along both the galaxy and the quasar sightline at the same moderate spectral resolution of MUSE. 
For the quasar sightline, we can use the high spectral resolution UVES spectrum ($8\,\kms$) to study the kinematics in more detail.  
In \Fig{fig:uves_spec}, we show one line each for \MnII{},
\ZnII{}, \FeII{}, \MgII{}, \MgI{}, \NaI{}. This is a subset of the low
ionization lines covered by the UVES spectrum.
In addition to the data, a multi-component fit is shown. For this fit, the positions and total number of velocity components in the absorption system were derived from all identified species. Their wavelength positions were then fixed to avoid degeneracy with blended features. For individual elements, only a subset of components was selected and fitted with a single Gaussian each with the evolutionary algorithm described in \citet{Quast:2005a} and applied in \citet{Wendt:2012a}. 

As expected from the MUSE spectrum, the $\MgII\,\lambda 2796$ absorption covers a broad velocity range - from $-130$ to $205\,\kms$ - and is strongly saturated for most of this range.
Unsaturated or weakly saturated lines, such as  the $\MgI\,\lambda2852$ line, are more useful to identify sub-structures.
Based on these transitions, we identified three main components, which are indicated in \Fig{fig:uves_spec} and labeled with \compa{} (\colcompa{}), \compb{} (\colcompb{}), and \compc{} (\colcompc{}).
They are offset from the systemic redshift of the foreground galaxy by $\vuvescompa$, $\vuvescompb$, and $\vuvescompc\,\kms$, respectively.

From the UVES spectrum, [\ZnII{}/\FeII{}] is measured for components A$+$B  to be $\sim1.1\pm0.1$,\footnote{The assumed solar abundances are adopted from \citealt{Jenkins:2009a} (based on \citealt{Lodders:2003a}).} which indicates
a significant amount of depletion for intervening systems \citep{DeCia:2016a} of
$\approx 0.3$~dex ($\approx 1.5$~dex) for Zn (Fe), respectively.
This level of depletion is also associated with more metal rich absorption systems
with [Zn/H] around $1/2$ solar \citep{DeCia:2016a}.

\begin{table}
  \caption{\label{tab:orientation} Geometrical orientation of the system.
    (1) Background object ID;
    (2) redshift;
    (3) impact parameter measured from \main{} $z=0.70$ foreground galaxy [$\kpc$ at redshift of \main{}];
    (4) azimuthal angle w.r.t. the major axis of \main{} [deg];
    (5) magnitude in 1\arcsec{} diameter aperture measured in pseudo-V filter created from MUSE data.
    }
  \centering
  \begin{tabular}{clrrrr}
    \hline
    Object          & z     & b    & $\alpha$ & $m_\textnormal{V}$ \\ %
    (1)             & (2)   & (3)  & (4)  & (5)   \\ %
    \hline
    Quasar          & 1.48  & $\bkpcquasarmm\pm0.7$ &  $\alphaquasarmm\pm8$  & 18.5  \\ %
    Back1           & 1.417 & $\bkpcbackonemm\pm0.7$  & $\alphabackonemm\pm9$ & 24.7 \\ %
    Back2           & 0.809 & $\bkpcbacktwomm\pm0.7$ & $\alphabacktwomm\pm8$ & 24.0 \\   %
    \hline
  \end{tabular}
\end{table}

\begin{table}
  \caption{\label{tab:prop_main_gal} 
  Physical properties of the foreground galaxy (\main{}). 
  For further details see \Sec{sec:gal_properties} and \paptwo{}.
  (1) \OII{} flux obtained from \gpk{} fit;
  (2) nebular extinction from $E(B-V)$-$M_*$ relation;
  (3) nebular extinction from SED fit;
  (4) instantaneous SFR from 1 \& 2;
  (5) instantaneous SFR from SED fit;
  (6) stellar mass from SED fit;
  (7) rest-frame B absolute magnitude from best fit SED model;
  (8) distance from the MS (assuming MS from \citealt{Boogaard:2018a});
  (9) age of galaxy from SED fit (time since onset of star-formation);
  (10) decay time in delayed $\tau$ SFH from SED fit;
  (11) rotation velocity from \gpk{} fit;
  (12) velocity dispersion from \gpk{} fit; 
  (13) virial velocity from $\vvir=\vmax/(1.1\pm0.3)$;
  (14) virial radius from $\vvir$;
  (15) virial mass from $\vvir$;
  (16) virial mass from abundance matching \citep{Behroozi:2010a};
  (17) escape velocities at position of quasar/\backone{} sightline assuming a truncated isothermal sphere.
  }
  \centering
  \begin{tabular}{r|r|r|r}
    \hline
  Row    & Property                    & Value                & Unit              \\
  \hline
  (1)  & $f_{\OII}$                  &   \fluxoiimain{}                   & $\uerglf$ \\
  (2)  & E(B-V) ($M_*$)   & \ebvmassmain{}       &    mag               \\
  (3)  & E(B-V) (SED)   & \ebvsedmain{}      &       mag            \\
  (4)  & SFR ($f_{\OII}$)  &       \sfroiiebvmassmain{}               & $\mpy$            \\
  (5)  & SFR (SED) &       \sfrinstsedmain{}               & $\mpy$            \\
  (6)  & $M_*$ (SED)                 & \mstarmain{}          & $\log(\msun)$     \\
  (7) & $B$ & \restbabsmain{} & mag \\
  (8)  & $\delta(MS)$                       &  \deltamsmain{}                    &   dex   \\
  (9)  & $age$                       & \agesedmain{}        & $\log(\mathrm{yr})$        \\
  (10)  & $\tau$                       &  \tausedmain{}                    & $\log(\mathrm{yr})$        \\
  (11)  & $\vmax$                     &  \vmaxmain{}                    & $\kms$            \\
  (12)  & $\sigma_{0}$                &  \sigmamain{}                    & $\kms$            \\
  (13)  & $\vvir$                     &  \vvirmain{}                    & $\kms$            \\
  (14)  & $\rvir$                     &  \rvirmain{}                    & $\kpc$            \\
  (15) & $\Mvir$ (from $M_*$) & \mvirfromstarsmain{} & $\log(\msun)$     \\ %
  (16) & $\Mvir$ (from kin.) &  \mvirfromkinmain{} & $\log(\msun)$     \\ %
  (17) & $\vesc$ (qso/\backone{}) & \vescqso{} / \vescbackone{} & $\kms$ \\
      \hline
    \end{tabular}
\end{table}

  \section{CGM toy model}
  \label{sec:model}

\MgII{} absorption around a galaxy is in observations predominantly found either along the galaxy's minor or major axis \citep[e.g.][]{Bordoloi:2011a, Bouche:2012a, Kacprzak:2012a, Nielsen:2015a, Martin:2019a}, see also \paptwo{} and \papthree{}. 
A natural explanation for this dichotomy is a simple model of a bi-conical outflow perpendicular to the galaxy disk and an extended gaseous disk aligned with the galaxy disk. 
This picture has gained support both from the theoretical and observational sides, i.e. predictions from cosmological hydro simulations (winds e.g.,~\citealt{Dubois:2008, Shen:2012a, ShenS:2013a}, disks: e.g.,~\citealt{Pichon:2011a,Kimm:2011a,ShenS:2013a,Danovich:2015a, Stewart:2011a,Stewart:2017a}) and directly observed emission properties of local galaxies (winds: e.g.,~\citealt{Veilleux:2005a} for a review, disks: e.g.,~\citealt{PutmanM:2009a,
  Wang:2016a, Ianjamasimanana:2018a}).

In the following, we investigate a toy model implementation for kinematics and
morphology of a disk$+$outflow model to interpret the observed absorption features in both the quasar and \backone{} sightlines. %

\subsection{Model parameters}
\begin{table*}
  \caption{\label{tab:toymodel:paraexplain}Summary of model parameters in the CGM toy model (cf. \Sec{sec:model}).}
  \centering
  
  \begin{tabular}{r|r|r|r}
    \hline
       & Property                    & Description                & Unit              \\
    \hline
    \multicolumn{4}{c}{Sightline} \\
    \hline
    (1) & $i$ &  Inclination  & [deg] \\
    \hline
    \multicolumn{4}{c}{Biconocial outflow} \\
    \hline
    (2) & $\thetaout$ & Outer (half-)cone opening angle  & [deg] \\
    (3) & $\thetain$ & Inner (half-)cone opening angle  & [deg] \\
    (4) & $\vout$ & Outflow velocity & [$\kms$] \\
    (5) & $\siggas$ & Gas velocity dispersion & [$\kms$] \\
    (6) & $\rho_1$ & Density at norm radius ($\MgII$) &  [$\mathrm{cm}^{-3}$] \\
    \hline
    \multicolumn{4}{c}{Extended gas disk} \\
    \hline
    (7) & $\vcirc$ & Circular velocity of gas & [$\kms$] \\
    (8) & $\vr$ & Radial velocity of gas & [$\kms$] \\
    (9) & $\hr$ & Exponential scale length (radial)& [kpc] \\
    (10) & $\hz$ & Exponential scale length (vertical)& [kpc] \\
    (11) & $\siggas$ & Gas velocity dispersion & [$\kms$] \\
    (12) & $\rho_0$ & Density at $r=0$ and $z=0$ ($\MgII$) &  [$\mathrm{cm}^{-3}$] \\
    \hline
    \end{tabular}

  \end{table*}

\subsubsection{Biconical outflow}
\label{sec:model:biconicaloutflow}

For the outflow model, we assume that a galaxy launches winds from its central region  into a bi-conical outflow  with half-opening angle $\thetaout$.  We allow the cone to be devoid of $\MgII$ within an inner opening angle, $\thetain$, as
indicated by larger samples of wind pairs  (e.g.~papers I\,\&\,III, \citet{Bouche:2012a}).
For the wind kinematics, we assume that the gas flows outward radially with an outflow velocity, $\vout$, that does not change with distance from the galaxy. From mass conservation, this constant velocity necessitates a radial density  $\rho(r)\propto r^{-2}$, which is normalized at 1~kpc with $\rho_1\equiv\rho(1\hbox{kpc})$.
We also account for random motions of the encountered gas with $\siggas$.
Moreover, we assume that the gas does not change its ionization state and that it is smoothly distributed.
Thus the wind parameters are $\thetaout$, $\thetain$, $\vout$ and $\rho_1$ and $\siggas$ which are listed Table~\ref{tab:toymodel:paraexplain}.  

The cone opening angle $\thetaout$ is   $\approx30\deg$, and the inner cone is $\thetain\approx15\deg$,  consistent with typical values in \papthree{}.
The outflow velocity $\vout$ is assumed  to be $150\,\kms$, corresponding to the typical $\vout$ in \papthree{}.
The intrinsic dispersion $\siggas$ is chosen somewhat arbitrarily to be $10\,\kms$.
All parameters of the fiducial model are summarized in Table~\ref{table:models:params}.

\subsubsection{Extended gas disk}
\label{sec:model:disk}

 However, as the sightlines are at relatively small impact parameters (at $\bkpcbackonemm\,\kpc$ and $\bkpcquasarmm\,\kpc$), a contribution from a thick extended gas disk cannot be ruled out. We model this extended gaseous disk as an exponential profile with scale length $\hr$ in radial direction.
In the direction perpendicular to the disk (z-direction), we assume an {exponentional profile with scale height $\hz$.
The gas density is normalized at the disk mid-plane in the disk center with $\rho_0$.}
For the disk's kinematics, we assume that the gas is rotating parallel to the
disk midplane with a circular velocity $\vcirc$, which we assume to be identical to $\vmax$ from the galaxy rotation. %
In addition, the gas velocity vector can also have a radial infall component, $\vr$, which is added to the tangential component keeping $\vcirc$ constant.~\footnote{The circular and the radial moving gas are here asssumed to add to a single components as in \paptwo{, but unlike in \citet{Bouche:2016a}, where the same gas has both a radial and infalling component.}}
The disk parameters are $\vcirc$, $\vr$, $\siggas$, $\hz$ and $\rho_0$  which are summarized in Table~\ref{tab:toymodel:paraexplain}.  

The circular velocity $\vcirc$ is given by the kinematics of the host galaxy as described in \S~\ref{sec:gal_properties}. 
The stellar scale height $h_z$ of distant galaxies is typically $1\,kpc$, as suggested by studies of edge-on disks in {\it Hubble} deep fields \citep[e.g.][]{Elmegreen:2006a,Elmegreen:2017a}.
We assume that the extended cool gas disk probed by \MgII{} has similar scale height ($h_z=1\,\kpc$). %
The gas dispersion, $\siggas$, is assumed to be $\sim10$~\kms\ appropriate for the temperature of low-ionization gas.
 The density $\rho_0$ will be adjusted in order to match the absorption optical
 depth for \MgI{}.

\subsection{Simulated absorption lines}

We use our code \cgmpy{} to calculate the \MgII{} absorption profile which the outflow cones  and/or the extended gas disk would imprint on a background source.
In short, the code calculates for each of small steps ($=1\,\mathrm{pc}$) along the line-of-sight (LOS) the LOS velocity, $v_\mathrm{los;step}$, and the column density, $N_\mathrm{step}$, which can subsequently be converted to an optical depth, $\tau_{\mathrm{step}}(\vlos)$.
The full $\tau(\vlos)$ distribution for the complete sightline is then obtained
by summing up the $\tau_{\mathrm{step}}(\vlos)$ from each step and each
component without the turbulent velocity dispersion $\siggas$.
We account for this random motions of the gas  ($\siggas$)  
by convolving the optical depth distribution with a Gaussian of the selected $\siggas$.
Finally, the absorption profile is obtained by taking $e^{-\tau(\vlos)}$ and convolving with the instrumental line spread function (LSF).

In the case of an extended sightline (such as for `\backone{'), the absorption
from the extended object is calculated by taking the average over individual
sightlines flux weighted over an elliptical aperture centered on the galaxy (for
\backone{} with an area of $\sim1\,\mathrm{arcsec}^2$).}

\section{Discussion}
\label{sec:discussion}

  \begin{table*}
    \caption{\label{table:models:params}
    The choice for each of the parameters in \Tab{tab:toymodel:paraexplain} as used for the five models described in \Sec{sec:discussion} and shown in \Fig{fig:model_comparison}. \newline
  }
    \begin{tabular}{r | cccccccccccc}
      \hline
      \textbf{Model} & 
      $i$ & $\thetaout$ & $\thetain$  & $\vout$ &  $\siggas$ &  $\rho_1$ &  $\vcirc$ & $\vr$ &  $\hr$ & $\hz$ & $\siggas$ & $\rho_0$ \\
      & (1) & (2) & (3) & (4) & (5) & (6) & (7) & (8) & (9) & (10) & (11) & (12) \\
      \hline
     Fiducial wind  &  71 & 35 & 15 & 150 & 10   & $8 \times 10^{-5}$ & -- & -- & -- & -- & --  & --  \\
     Slow wind    &  71 & 35 & 15 & 75 & 10     & $8 \times 10^{-5}$ & -- & -- & -- & -- & --  & -- \\
     Disk  &  71 & -- & -- & --  & -- & -- & 118 & --  & 5 & 1 &  10  &  $3\times10^{-3}$\\
     Disk w. infall &  71 & -- & -- & --  & -- & -- & 118 & -40  & 5 & 1 &  10  & $3\times10^{-3}$ \\
     Disk + wind  &  71 & 35 & 15 & 100  & 10 & $8 \times 10^{-5}$  & 118 & --  & 5 &1 &  10  & $3\times10^{-3}$ \\
     \hline
   \end{tabular}
\end{table*}

\begin{figure*}
\includegraphics[width=\textwidth]{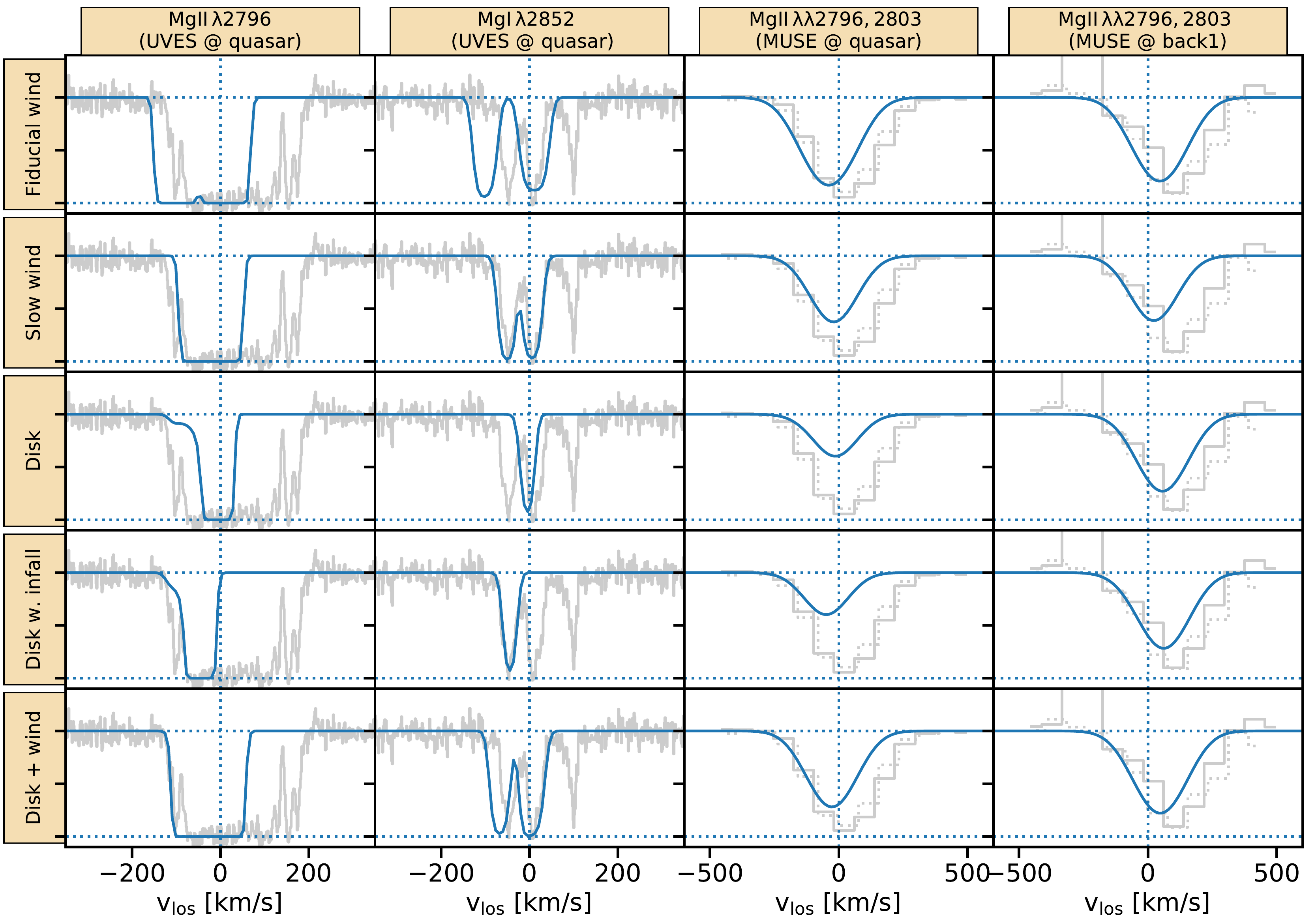}
\caption{\label{fig:model_comparison} 
Comparison between data and various models. Eeach row shows a different model, as listed in \Tab{table:models:params}.
In each panel the light grey curve is the observed  absorption, while the blue line shows the modeled absorption. 
The first two columns show \MgII{}$\,\lambda2796$ (column 1) and \MgI{}$\,\lambda 2852$ (column 2) in the quasar sightline at the resolution of UVES.
The third and fourth column show \MgII{}$\,\lambda2796, 2803$ for the quasar sightline (column 3) and the \backone{} galaxy sightline (column 4) at the resolution of MUSE. Here, the solid line is \MgII{}$\,\lambda2796$ (data and model) and the dotted line is \MgII{}$\,\lambda2803$ (data). 
}
\end{figure*}

Here, we describe how the toy model discussed in  \Sec{sec:model} performs in discribing our data. However, we stress that we do not expect this simple toy model to account for all data features nor do we attempt to formally fit it to the data.
Thus, if the model can, at least approximately, explain most of the absorption in \emph{both} background sightlines, the simple toy model can be viewed as a  description of the $\main$ galaxy's CGM.

\subsection{The fiducial (wind-only) model}
\label{sec:discussion:wind}

  We first tested the performance of a fiducial biconical outflow-only model (cf.~\Sec{sec:model:biconicaloutflow})  given that both the quasar and \backone{} are positionned along the minor axis of the host galaxy, i.e. without an extended gas disk.

Here, the model's orientation is set by the measurement of the galaxy's inclination $i$ (see \Sec{sec:gal_properties}). 
However, as the sign of the galaxy inclination cannot be constrained with the
available data \citep[see e.g.][]{Ho:2019a}, we were left with two possible solutions.
Here, we choose the sign of the inclination such that the absorbing gas in the cones is outflowing. 
This outflow assumption requires that redshifted absorption must originate from the far-side cone, and consequently, the \backone{} galaxy sightline crosses this far-side cone.
Panels B and C of \Fig{fig:kinematics_model} show the adopted orientation.

For our `fiducial' outflow model, we assume a value for $\thetaout$ ($35\deg$),
which is at the higher end of typical values found in \papthree{}. We made this
choice, to ensure very high coverage over the extended \backone{} galaxy
sightline in the model, as required by the observed absorption strength (see
\Sec{sec:mgii_abs_muse}).

In  \Fig{fig:model_comparison} (row~1 - `Fiducial wind'), we overlay the resulting absorption profiles over the UVES and MUSE data for the `qso' (Cols 1, 2 and 3) and `\backone' (Col. 4) sightlines.  
Column 1 (2) show the  model for the quasar sightlines  for \MgII{} (\MgI),
respectively, where we scaled the \MgI{} density by 1/600 compared to \MgII{}
according to \citet{Lan:2017a}.
Comparing our UVES data to the model for the quasar sightlines shown in Cols.~1 ans 2,
   we find that the absorption is made of two separate components which arise from the assumption of an empty inner cone. 
These two components might correspond to components \emph{A} and \emph{B} in the observed spectrum (see \Sec{sec:uves_abs}).
 Comparing our MUSE data and the fiducial wind model (Cols 3 and 4), we find that the model and data match qualitatively for the blue- (red-) shifted absorptions  in the quasar (galaxy) sightlines absorption shown in Col.~3 (4), respectively.
However, there are some discrepancies between the model and the data.

The main discrepancy is that the wind model cannot explain the redshifted third component C.
Another discrepancy is that, for the quasar absorption, the model predicts a blue-shift ($-75$~\kms)
whereas the observed absorption is close to systemic at $\approx+25$~\kms.

A model with lower outflow velocity ($\vout\approx 75\,\kms$) would better match to components A and B in the \MgI{} absorption (\Fig{fig:model_comparison}; row~2 - `Slow wind'). However, it under-predicts the redshift compared to the \MgII{} data in the \backone{} galaxy sightline.
Note that this potential velocity difference between the two sightlines could indicate deceleration of the gas with distance from the galaxy, as the quasar sightline is probing gas at a larger impact parameter than the \backone{} sightline does ($\bkpcquasarmm\,\kpc$ vs $\bkpcbackonemm\,\kpc$). Strong, non-gravitational, deceleration in an outflow could be  due to drag forces (in observations e.g., \citealt{Martini:2018a}; in simulations e.g.,~\citealt{Oppenheimer:2010a}).
 However, this interpretation would require the strong assumption that the two opposite cones have the same velocity profile.

\subsection{{Disk model}}
\label{sec:discussion:disk}

Given the limitations of the fiducial wind only-model, and the relatively small impact parameters, we discuss the extended gaseous disk model presented in \Sec{sec:model:disk}. 
Indeed, the two minor-axis sightlines cross the disk midplane at galactocentric radii of $\Rgalkpcbackone$ ($\Rgalrvirbackone$) and $\Rgalkpcqso$ ($\Rgalrvirqso$), within the extent of co-rotating gas disks from  \paptwo{} and \citet{Ho:2017a}.
Before discussing a potential combination of wind and disk-model, we test whether a simple thick disk model similar to \citet{Steidel:2002a, Kacprzak:2010a, Ho:2017a} can potentially explain all absorption on its own.

In  \Fig{fig:model_comparison} (row~3 - `Disk'), we overlay the resulting absorption profiles over the MUSE and UVES data as before.
Comparing the UVES data to our model shows that a thick disk model can only
explain component B in the \MgI{} spectrum.\footnote{
We note that a very thin disk   would have a narrower profile, hence a lower equivalent width, and also a lower velocity shift than a thick disk.  This is, because a sightline crossing a thick disks encounters different velocities at different heights above the disk, up to $\sin(i)\vcirc$ \citep[e.g.][]{Steidel:2002a}.}
As for the wind model, the thick disk model cannot explain the redshifted third component C.
However,  component A in the UVES spectrum could be accounted for with an extension of this disk model with a radial inflow component (shown in row 4 of \Fig{fig:model_comparison} - `Disk + infall').
The observed velocity of $\vuvescompa\,\kms$ would require a radial velocity
component of
$\vr\approx\vrforcompAkms=\vrforcompAvvir$.\footnote{$\vr\approx\vrforcompAkms$ is
  enough to match the observed blueshift of component A, because the model has also a
  contribution from the rotational component ($\vcirc$).} Such a radial inflow velocity is
feasible, based on results from simulations \citep[e.g.][]{Rosdahl:2012a, vandeVoort:2012a, Goerdt:2015a, Ho:2019b} and observational studies
targeting the major axis sightlines (e.g.~\citealt{Bouche:2013a,  Bouche:2016a, Rahmani:2018a}, \citetalias{Zabl:2019a}).

 \subsection{Combined disk and wind}
 \label{sec:discussion:disk_and_wind}

The observed absorption might be a combination of aborption from both a disk and an outflow component.
As discussed in \Sec{sec:discussion:wind}, the ouflow component alone, a faster wind ($150\,\kms$) matches better the observed absorption in the \backone{} sightline, while
a slower wind ($75\,\kms$) matches better the absorption in the qso sightline.
For the following, we assume a wind speed of $100\,\kms$ as a compromise to match approximately both sightlines with a single wind speed.  \Fig{fig:model_comparison} (row 5 - `Disk + wind') shows the resulting absorption profile when combining the disk and this wind toy model (The same model is also shown in \Fig{fig:kinematics_model}, panel D).
While imperfect, the toy model is qualitatively in agreement with the observed spectra, apart from component C.
Component C  might be an unrelated component, similar to the high-velocity clouds (HVC) seen around the Milky Way (e.g.,~\citealt{Wakker:1997} for a review).
In summary, a plausible interpretation of the observed kinematics in the two sightlines is absorption in a bi-conical outflow with a potential disk contribution.

\subsection{Feasibility of the outflow} 

\label{sec:outflow_feasibility}

As discussed in \Sec{sec:gal_properties}, the $\mbox{SFR}$ of the \main{} galaxy is low compared to star-forming galaxies with similar mass at similar redshift.
This raises the question whether the energy and the momentum that are required to explain the wind are at all feasible.
To answer this question, we estimated the mass outflow rate, $\mdotout$, the energy-outflow rate, $\Edotout$, and the momentum outflow rate, $\pdotout$.
These estimates can subsequently be compared to the estimated $\mbox{SFR}$ and the
corresponding energy and momentum deposition rates from supernovae (SNe).

We estimated $\mdotout$ for the bi-conical outflow of cool gas using Eq.~5 from \papthree{}.
As inputs to the equation we assumed $\thetaout=35\deg$, $\thetain=15\deg$, $b=15\,\kpc$, $\vout=100\,\kms$, $\log(N_{\HI}/\ucoldens)=20.0$.
Here, we estimated the $\HI$ column density using the $\rewmgii$ - $\HI$ relation from \citet{Menard:2009a} and \citet{Lan:2017a}, which has an uncertainty of around $0.3\,\dex$.
Using these values in the equations we obtain $\mdotout=2.0\,\mpy$. This corresponds with the assumed $\vout=100\,\kms$ to $\Edotout=6.0\times10^{39}\,\ue$ and $\pdotout=1.3\times10^{33}\,\um$. 

A comparison of $\mdotout$ to the estimated $\mbox{SFR}$ allows us to infer the mass-loading ($\eta=\mdotout/\mbox{SFR}$), which characterizes the efficiency of a star formation powered wind to remove gas from the galaxy.
Assuming that the wind was powered by the current SFR of $0.5\,\mpy$, we infer $\eta\approx4$. 
This value can be compared to measurements of $\eta$ both from individual estimates (quasar sightlines e.g.~\papthree{}, \citealt{Bouche:2012a, Schroetter:2015a}; down the barrel: e.g. \citealt{Weiner:2009a,Martin:2012a,Rubin:2014a, Sugahara:2017a}), indirect observational evidence \citep[e.g.][]{Zahid:2014a, Mitra:2015a}, or simulations \citep[e.g.][]{Hopkins:2012a, Muratov:2015a}.
For the mass and redshift of our main galaxy, the values in these studies typically range  from $\eta \approx 1\mbox{--}10$ (see also discussion in \papthree{}). 
Hence, we conclude that the $\eta$ corresponding to our preferred model seems feasible.

A direct comparison of the measured $\Edotout$ and $\pdotout$ to the momentum and energy injected by SNe leads to a similar conclusion. 
Per $1\mpy$ of star formation SNe deposit mechanical energy and momentum  with
rates of approximately $1.6\times3\times10^{41}\,\ue$ (from
\citealt{Chisholm:2017a} based on \citealt{Leitherer:1999a}) and $1.6\times
2\times10^{33}\,\um$ (\citealt{Murray:2005a}).\footnote{Factor 1.6 is to convert from
  the \citet{Salpeter:1955a} to the \citet{Chabrier:2003a} IMF.}
This means that our measured values correspond to energy and momentum loading of 3\% and 80\%, respectively. 
These values are comparable to those found by \citet{Chisholm:2017a} for a sample of local star-forming galaxies when considering the relevant mass range.\footnote{We have only included the cool phase of the outflow, so the total loading factors could be higher.}

Finally, we note that the actual loading factors could be smaller. The SFR might have been higher at the time when the wind was launched.
It would have taken the wind $\approx 200\,\myr$ ($\approx 100\,\myr$) to travel to the quasar (\backone{}) sightline, assuming $\vout=100\,\kms$. 
With the limited available data we cannot rule out that there was a significant burst of star-formation about $200\mbox{--}300\,\myr$ ago, as motivated by tests with non-parmetric SFHs with \ppxf{} \citep{Capellari:2004a, Capellari:2017a}.

\section{Conclusions}
\label{sec:conclusions}

It is now statistically well established that there is a dichotomy in the
spatial distribution of the cool circum-galactic medium (CGM) gas probed through
\MgII{} absorption, where the two components have been identified as arising in an extended
gas disk and a bi-conical outflow.
In this paper, we present a rare chance alignment of a quasar and a UV-bright
background galaxy at relatively small impact parameters ($\bkpcquasarmm$
and $\bkpcbackonemm \,
\kpc$) from a $z=0.7$ foreground galaxy. 
As the two sightlines are close to the foreground galaxy's projected minor axis,
but on opposite sides of the major axis, the configuration is ideal to test the
bi-conical outflow component.
Through studying the observed absorption both in MUSE and UVES data from the
\mfl{} survey, and comparison to modeled absorption, we reached the following conclusions:

\begin{itemize}
  \item Both sightlines show very strong \MgII{} absorption ($\rewmgii>2.0\,\textnormal{\AA}$). 
  \item We find a significant velocity shift of $84\pm17\,\kms$ between the two sightlines.
  \item The observed velocity shift is in broad agreement with
    a  bi-conical outflow toy model with a moderate outflow velocity of $\approx 100\,\kms$, possibly combined with
    a disk model.
  \item The foreground galaxy has a relatively low $\mbox{sSFR}$ ($\ssfroiiebvmassmain\,\mathrm{Gyr}^{-1}$), which puts the galaxy  $0.6\,\dex$ below the MS at $z=0.7$. However, the mass-loading ($\eta$) required to explain the modelled outflow is not unrealistic high ($\eta\approx4$). Moreover, the sSFR may have been higher when the wind was launched, $\sim 10^8\,\mathrm{yr}$.
\end{itemize}

This study presented a `tomographic' study (i.e. with multi-sightline) of the CGM 
around an individual galaxy in the distant Universe ($z\approx0.7$), and hence goes beyond the
statistical inference from single sightline samples.
While we find the data to be in broad agreement with our fiducial CGM model, we cannot rule out alternative explanations. 
A comparison of the CGM model to larger samples of rare multi-sightline cases, including cases with even more sightlines as e.g. provided by background groups or gravitationally lensed arcs \citep[e.g.][]{Lopez:2018a, Lopez:2019a}, will be an important test for our assumed geometry.
Additionally, it will be necessary to test the geometry against observations of the CGM in emission \citep[e.g.][]{Finley:2017a, Rupke:2019a}.

\section*{Acknowledgements}

This study is based on observations collected at the European Southern
Observatory under ESO programmes 097.A-0138(A), 097.A-0144(A), 0100.A-0089(A). 
This work has been carried out thanks to the support of the ANR FOGHAR (ANR-13-BS05-0010), the ANR 3DGasFlows (ANR-17-CE31-0017), the OCEVU Labex (ANR-11-LABX-0060), and the A*MIDEX project
(ANR-11-IDEX-0001-02) funded by the ``Investissements d'avenir'' French
government program.

This  work  made  use  of  the  following  open  source
software:
 \textsc{GalPak3D} \citep{Bouche:2015a},
 \textsc{ZAP} \citep{SotoK:2016b},
  \textsc{MPDAF} \citep{Piqueras:2017a},
  \textsc{matplotlib} \citep{Hunter:2007a},
  \textsc{NumPy} \citep{vandderWalt:2011a, Oliphant:2007},
  \textsc{Astropy} \citep{Robitaille:2013a}.

\bibliographystyle{mnras}
\bibliography{main_jz.bib} %

\appendix
\section{Uncertainty on inclination and position angle}
\label{app:uncert_incl_posangl}

\begin{figure*}
  \includegraphics[width=0.7\textwidth]{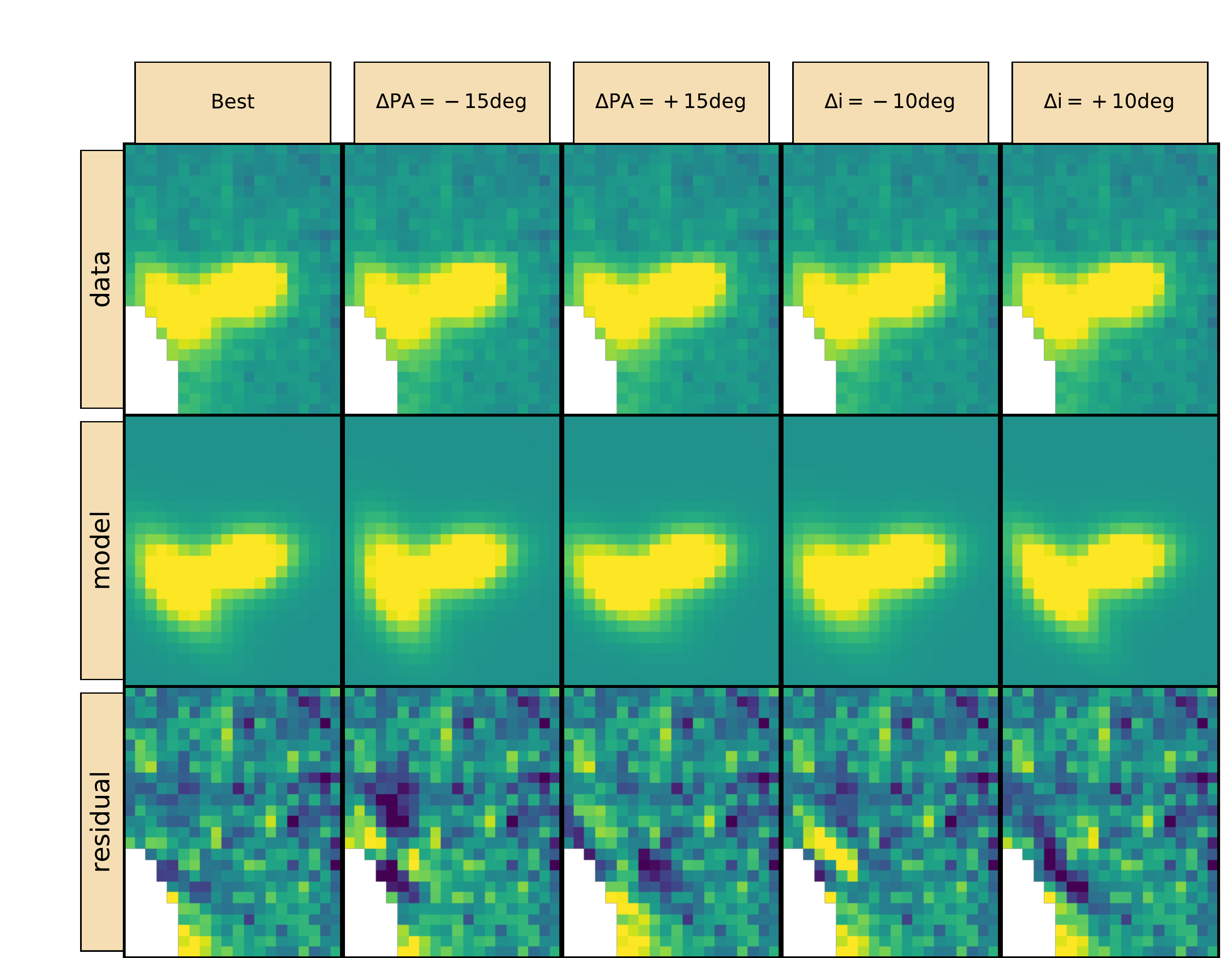}
  \caption{\label{fig:app:uncert_incl_posangl}Data, \galfit{} model, and residuals
    (data-model) are shown for each of five models in the top, center, and
    bottom row, respectively.
    The data, which are identical in each of the four columns, are a pseudo
    broadband r image created from the MUSE cube. The
    \main{} foreground galaxy is to the left and the \backone{} background
    galaxy is to the right (north is to the top, east to the left; different
    orientation from \Fig{fig:kinematics_model}). The white region to the lower left masks
    residuals from the quasar subtraction.
    \textbf{Left column: } Best fit \galfit{} model, where both \main{} and \backone{}
    were fit simultaneously. The \main{} galaxy has best fit $PA_\mathrm{best}=37\deg$
    and $\incl_\mathrm{best}=71\deg$ assuming a $n=1$ Sersic profile.
    \textbf{Center left column:} This column and the other three columns show
    the best fit model with either the $PA$ or $\incl$ of the \main{} galaxy
    adjusted. Here, $PA=PA_\mathrm{best} - 15\deg$; 
    \textbf{Center column:}  $PA=PA_\mathrm{best} + 15\deg$; 
    \textbf{Center right column:}  $\incl=\incl_\mathrm{best} - 10\deg$;
    \textbf{Right column:}  $\incl=\incl_\mathrm{best} + 10\deg$.
  }
\end{figure*}

In our analysis, we tied the orientation of our toy model (\Sec{sec:model}) to the orientation of the \main{} foreground galaxy.
Therefore, a robust measurement of position angle ($PA$) and inclination ($\incl$) is important.
As discussed in \Sec{sec:gal_properties}, the measurement of the galaxy's morphology is somewhat complicated by residuals from the PSF subtraction.
The residuals made a formal assessment of the uncertainties based on the $\chi^2$ doubtful.
Therefore, we preferred to rely on a visual assessment of the uncertainties.
For this purpose we created \galfit{} models deviating from the best fit model either in $PA$  or inclination.
\Fig{fig:app:uncert_incl_posangl} shows models and residuals all for the best-fit model, the $PA_\mathrm{best}-15\deg$, $PA_\mathrm{best}+15\deg$, $\incl_\mathrm{best}-10\deg{}$, $\incl_\mathrm{best}+10\deg{}$.
Except for the modified $PA$ or $\incl$, we used in each case identical morphological parameters to those in the best fit model. 
The only free fit parameter in each of the alternative models was the total flux.
Both for $PA\pm15\deg$ and $\incl\pm10\deg$ the residuals are much stronger than for the best-fit model and the models seems essentially inconsistent with the data.
Therefore, it seems plausible to define these $PA$ and $incl$ differences as $2\sigma$ uncertainties.
In summary, we conclude therefore that the $1\sigma$ uncertainties for $PA$ and $\incl$ are $8\deg$ and $5\deg$, respectively.

In addition to the uncertainty in $PA$ and $\incl$, there is also a small
uncertainty on the centroid. We estimated this uncertainty through comparison
between the continuum centroid obtained from this \galfit{} fit and the \OII{}
centroid obtained from \gpk{} fit. We find a deviation of $0\arcsec.16$ between
the two centroids. Therefore, we can assume as $1\sigma$ uncertainty 
$0\arcsec.1$ both in right ascension and declination.

\section{Impact of uncertainties on inclination and position angle on models}
\label{app:impact_of_uncert}
\begin{figure*}
  \includegraphics[width=\textwidth]{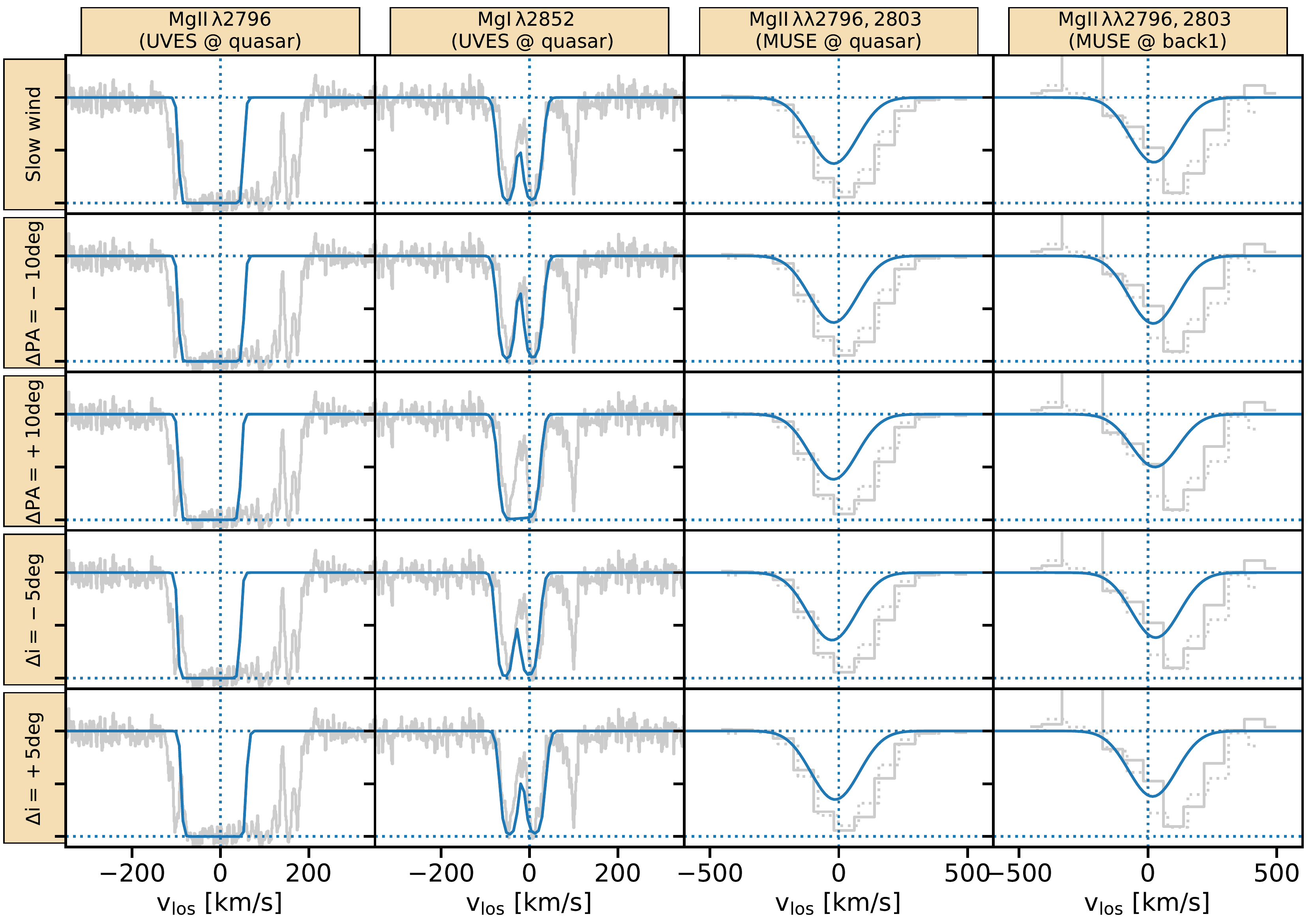}
  \caption{\label{fig:app:impact_of_uncert_wind}Comparison between modeled and observed absorption for the `Slow
wind' model assuming different inclinations and position angles. The first
row is identical to row 2 in \Fig{fig:model_comparison}, where the best fit
$\incl$ and $PA$ were assumed. Details about the content displayed in the four
columns are given in the caption of \Fig{fig:model_comparison}. The subsequent
rows (1-4) show the same model, but with $\incl$ or $PA$ changed by the
values stated in the row labels. For further details, see Appendix~\ref{app:impact_of_uncert}.}
\end{figure*}

\begin{figure*}
  \includegraphics[width=\textwidth]{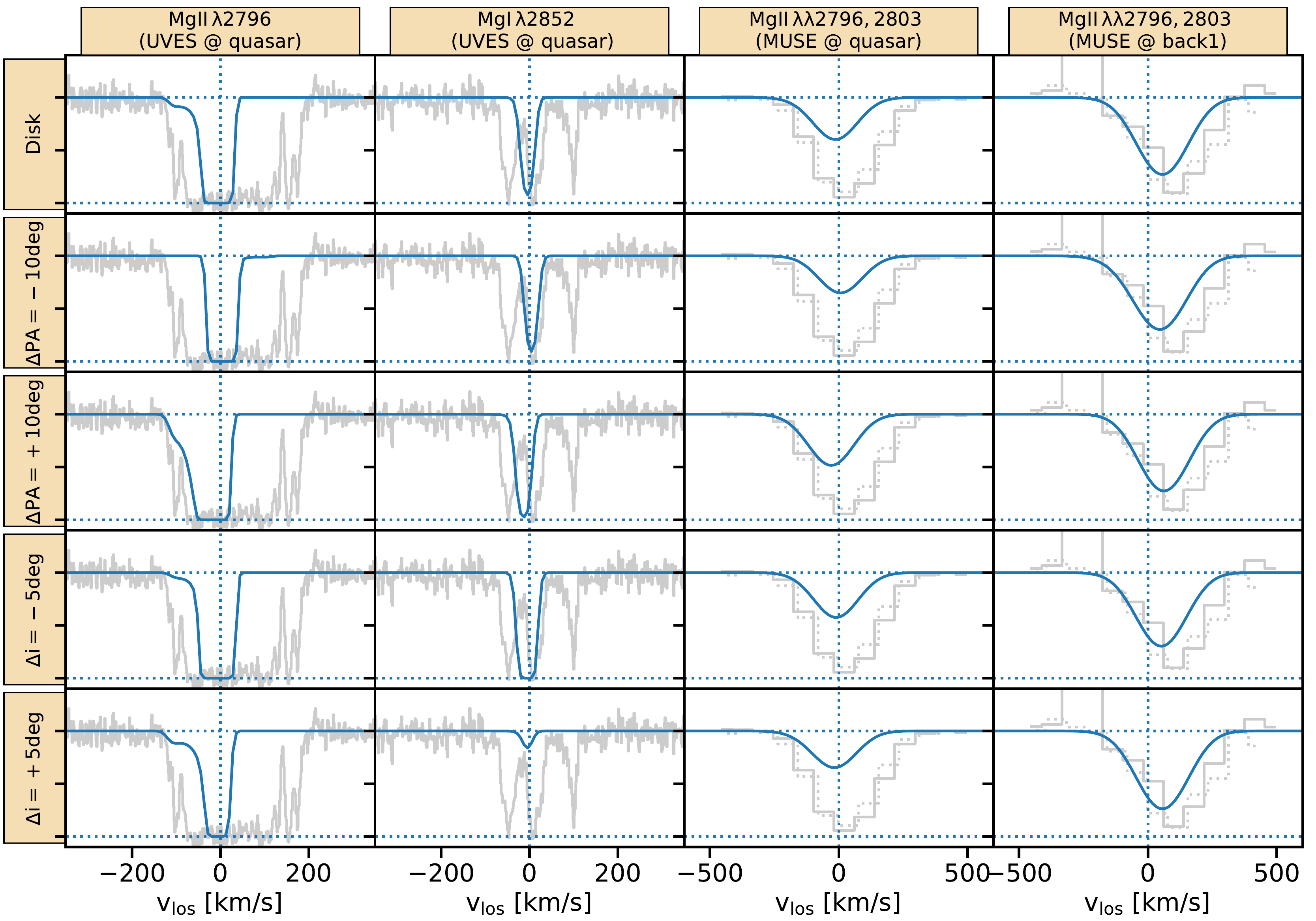}
  \caption{\label{fig:app:impact_of_uncert_disk} As \Fig{fig:app:impact_of_uncert_wind}, but here for the `Disk' model (cf.~row 3 in \Fig{fig:model_comparison}). For further details, see Appendix~\ref{app:impact_of_uncert}.}
\end{figure*}

In this section, we asses the impact of the uncertainties for $i$ and $PA$ on the simulated
absorption in our toy models.

In \Fig{fig:app:impact_of_uncert_wind} we show the `Slow wind' model
(see.~\Tab{table:models:params}) with either $\incl$ or $PA$ changed compared to
the fiducial values (row 1). Rows 2 and 3 show the result for changing the
$\incl$ by $\pm5\deg$ (i.e., $66\deg$ and $76\deg$), while keeping the fiducial
value for the $PA$. Rows 4 and 5 show the impact of varying the $PA$ of \main{}
by $\pm10\deg$. Assuming $\Delta PA\pm10\deg$ means that the azimuthal angle
$\alpha$ is changed by $\mp10\deg$ both for the quasar and the \backone{}
sightline (equally) compared to the values stated in \Tab{tab:orientation}. All
other parameters are kept identical to those listed for the `Slow wind' model in
\Tab{table:models:params} and shown in the first row of
\Fig{fig:app:impact_of_uncert_wind}.

In general, the differences between the absorption profiles for these variants
appear small. The strongest visible impact is for $\Delta PA = +10\deg$
(corresponding to $\alpha=71\deg$ for quasar and $\alpha=-119 \deg$ for
\backone{}). In this case, the \MgI{} absorption profile is not double-peaked
and the \MgII{} absorption in the \backone{} sightline is visibly weaker than in
the fiducial model. The double peak is absent, because the distance from the
minor axis is larger than in the fiducial case and, consequently, the quasar
sightline does not cross the hollow part of the cone. The weaker \MgII{}
absorption for \backone{} is also a consequence of a larger distance from the
minor axis. At $\alpha=-119 \deg$ part of the extended \backone{} galaxy
sightline is no longer covered by the cone at all, which reduces the effective
$\rewmgii$.

In Fig \ref{fig:app:impact_of_uncert_disk}, we test the impact of the same
$\incl$ and $PA$ variations, but now for the `Disk' model (see
\Tab{table:models:params}). Here, the differences in absorption strength appear
stronger than in the wind case. This is especially the case for changes in
$\incl$. Here, the strength varies - especially for the \MgI{} absorption in the
quasar sightline - as the sightline crosses the disk mid-plane at larger
galacto-centric radii, the larger the $\incl$ is. We note, though, that most of
the changes could be compensated for by merely choosing a disk with higher
density. For the variations with $PA$, the centroid of the absorption shifts, but only slightly.

In summary, we can conclude that the uncertainties on $\incl$ and $PA/\alpha$,
as estimated in Appendix \ref{app:uncert_incl_posangl}, only subtly change our
simulated profiles. Therefore, we can conclude that our conclusions in
\Sec{sec:discussion} are not impacted by these uncertainties.

\bsp	%
\label{lastpage}
\end{document}